\documentclass[sigconf,nonacm]{acmart}

\usepackage{booktabs}
\usepackage{makecell} % cell width
\usepackage{xcolor}
\AtBeginDocument{%
  }

\begin{document}

\title [NeuResonance] {NeuResonance: Exploring Feedback Experiences for Fostering the Inter-brain Synchronization}

\author{Jamie Ngoc Dinh}
\affiliation{%
  \institution{University of Maryland}
  \city{College Park}
  \country{USA}}
\email{ngocdinh@umd.edu}
\authornote{Equal Contribution.}

\author{Snehesh Shrestha}
\affiliation{%
  \institution{University of Maryland}
  \city{College Park}
  \country{USA}}
\email{snehesh@umd.edu}
\authornotemark[1] 

\author{You-Jin Kim}
\affiliation{%
  \institution{Texas A\&M University}
  \city{College Station}
  \country{USA}}
\email{yujnkm@tamu.edu}

\author{Jun Nishida}
\affiliation{%
  \institution{University of Maryland}
  \city{College Park}
  \country{USA}}
\email{jun@umd.edu}

\author{Myungin Lee}
\affiliation{%
  \institution{University of Maryland}
  \city{College Park}
  \country{USA}}
\email{myungin@umd.edu}

\renewcommand{\shortauthors}{Dinh et al.}

\begin{abstract}

When several individuals collaborate on a shared task, their brain activities often synchronize. This phenomenon, known as Inter-brain Synchronization (IBS), is notable for inducing prosocial outcomes such as enhanced interpersonal feelings, including closeness, trust, empathy, and more. Further strengthening the IBS with the aid of external feedback would be beneficial for scenarios where those prosocial feelings play a vital role in interpersonal communication, such as rehabilitation between a therapist and a patient, motor skill learning between a teacher and a student, and group performance art. This paper investigates whether visual, auditory, and haptic feedback of the IBS level can further enhance its intensity, offering design recommendations for feedback systems in IBS. We report findings when three different types of feedback were provided: IBS level feedback by means of on-body projection mapping, sonification using chords, and vibration bands attached to the wrist.

\medskip

\textit{This is a preprint version of this article. The final version of this paper can be found in the Proceedings of ACM CHI 2025. For citation, please refer to the published version. This work was initially made available on the author's personal website [yujnkm.com] in March 2025, and was uploaded to arXiv for broader accessibility. }

\end{abstract}

% 142 / 150 words max

%\textbf{The proposed work, \textit{NeuResonance} is an interactive multimodal system that visualizes multiple users' physiological activity, including brain activity recorded from electroencephalogram (EEG) and muscle activity recorded from electromyogram (EMG), on their own bodies by means of on-body projection mapping technique.}
%With this system, one user can visually observe another's brain and muscle activity simply by looking at the other user's body, which would potentially influence their own neural activities each other interactively.




\begin{teaserfigure}
  \includegraphics[width=1.0\textwidth]{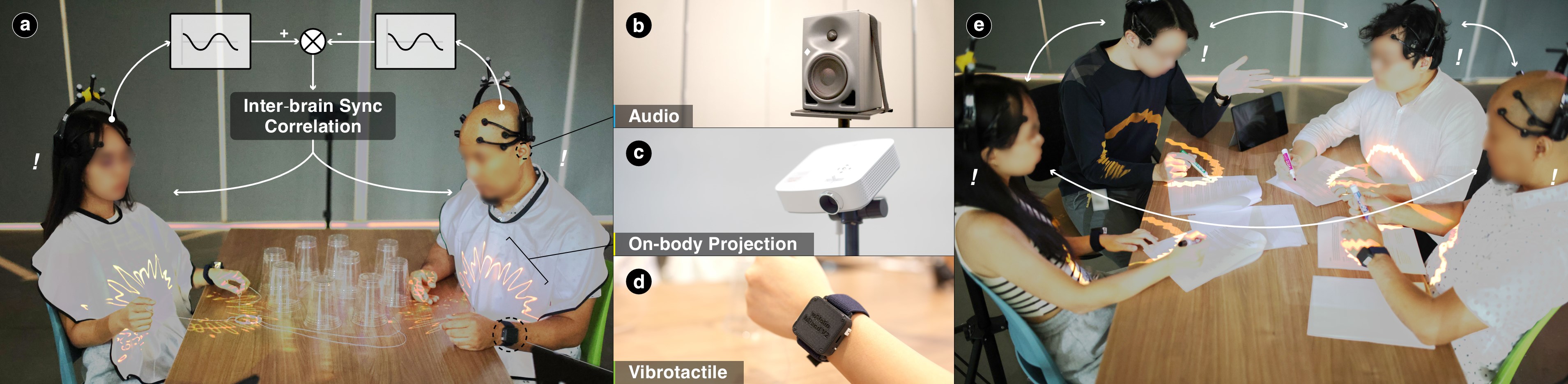}
  \caption{(a) When several individuals work together on a shared task, their brain activities tend to synchronize. This phenomenon, known as \textit{Inter-brain Synchronization} (IBS), is linked to positive social outcomes. We investigated whether providing feedback on its intensity through (b) auditory, (c) visual, and (d) haptic modalities could further amplify the effect of IBS. (e) This approach could potentially improve team communication and performance, for example, during group brainstorming sessions.}
  \Description{(a) When several individuals work together on a shared task, their brain activities tend to synchronize. This phenomenon, known as Inter-brain Synchronization (IBS), is linked to positive social outcomes. We investigated whether providing feedback on its intensity through (b) auditory, (c) visual, and (d) haptic modalities could further amplify the effect of IBS. (e) This approach could potentially improve team communication and performance, for example, during group brainstorming sessions.}
  \label{fig:rep}
\end{teaserfigure}

\maketitle

\section{Introduction}

\begin{figure*}[t]
    \centering
    \includegraphics[width=1.0\linewidth]{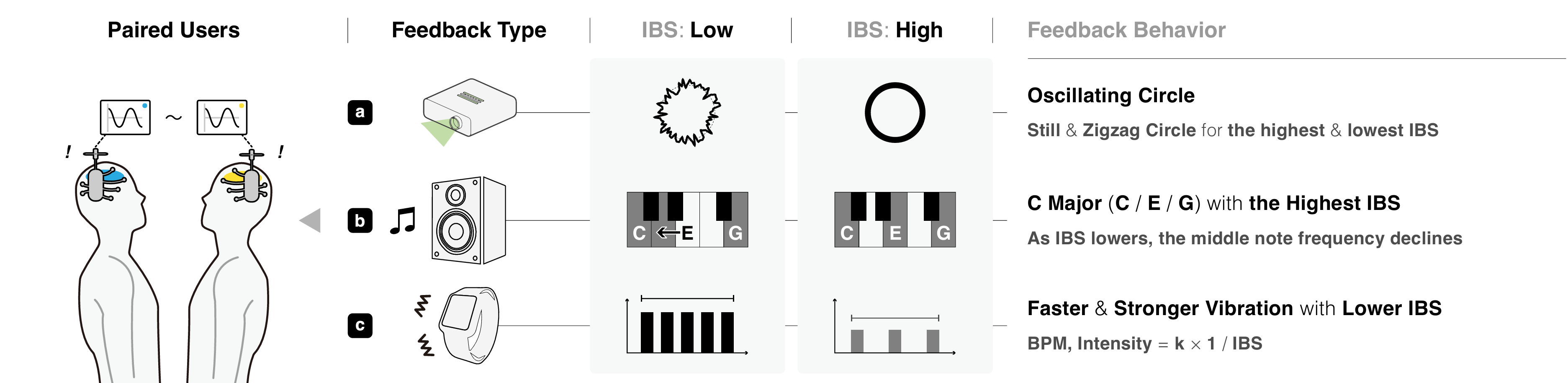}
    \caption{Design approaches for the (a) Visual, (b) Auditory, and (c) Haptic feedback}
    \Description{Design approaches for the (a) visual, (b) auditory, and (c) haptic feedback}
    \label{fig:approach}
\end{figure*}

When individuals perform tasks together, their brain activities can synchronize, a phenomenon known as Inter-brain Synchronization (IBS). This effect has been widely studied in neuroscience, especially in the context of multi-user interactions ~\cite{semertzidis_psinet_2024, hu_brain--brain_2017,will_brain_2007, shehata_team_2021, valencia_what_2020}. Studies have observed shared neural patterns during cooperative problem-solving and synchronized physical activities, including watching the same movie ~\cite{ahmadzadeh_does_2019}, performing coordinated gestures ~\cite{dumas_inter-brain_2010}, playing musical instruments together ~\cite{pai_neuraldrum_2020, sanger_intra-_2012}, engaging in online games ~\cite{wikstrom_inter-brain_2022}, holding hands ~\cite{goldstein_brain--brain_2018}, solving puzzles ~\cite{fishburn_putting_2018}, and typing passages collaboratively ~\cite{reinero_inter-brain_2021}.

%% Prosocial Outcomes 

Prior research has shown \textbf{the association of IBS with prosocial and interpersonal outcomes}. For example, Shiraishi et al. ~\cite{shiraishi_inter-brain_2021} found that IBS is closely linked to behavioral coordination, which enhances `we-mode' processing and fosters a stronger sense of joint agency among participants. Other studies have demonstrated correlations between IBS and feelings of closeness ~\cite{liu_team-work_2021}, trust ~\cite{liu_interactive_2018}, empathy ~\cite{goldstein_brain--brain_2018}, pain ~\cite{goldstein_brain--brain_2018}, agreeableness ~\cite{zhang_group_2021}, and more. Additionally, the IBS also has the potential to predict collaborative activities by refining the subjective aspects of interactions, including internal experiences and perceptions that individuals bring to and derive from their interactions with others ~\cite{reinero_inter-brain_2021}.

While studies have demonstrated the correlation between IBS and positive prosocial outcomes, research on methods to \textbf{actively enhance IBS remains unexplored}, possibly because IBS is still an emerging area in HCI ~\cite{semertzidis_psinet_2024, gumilar_comparative_2021}. Semertzidis et. al ~\cite{semertzidis_psinet_2024} developed a wearable brain-to-brain system to augment IBS through magnetic brain stimulation. Their study demonstrated that enhancing IBS fostered group empathy, group performance, and relational interactions.
While these findings highlight the potential of IBS in facilitating better prosocial experiences, little is known about optimal feedback design and efficacy for enhancing IBS through external feedback that is widely accessible to end-users.
%IBS itself in the study is not directly perceivable through visual contact, and the system does not leverage human-fundamental modalities. %not sure about the "but" part

%% Scenario example where the prosocial outcomes become beneficial
To address this gap in the literature, we propose a study aimed at answering the question: \textbf{How can external feedback be used to actively enhance IBS between individuals?} Our findings would serve as an initial step toward achieving unique teamwork and creating synergies beyond what individuals can achieve alone. For instance, enhancing IBS could be beneficial in educational settings where students collaborate on complex problems, in corporate environments to improve teamwork and communication, or in therapeutic contexts to strengthen the bond and sympathy between therapist and patient ~\cite{miskovic_changes_2011}. 

This study explores how different types of external feedback—visual, auditory, and haptic—can enhance IBS during collaborative tasks. The hypotheses (H1-H4) propose that these feedback modalities can increase IBS levels, with external feedback further augmenting synchronization. The study employs dual-EEG measurements using Emotiv EPOC X headsets that capture real-time brain activity from two participants. A motion capture system (Vicon) helps remove motion artifacts, and IBS level is calculated using the circular correlation coefficient (CCorr). The feedback system includes on-body projection mapping for visual feedback, frequency shifts in musical chords for auditory feedback, and vibrotactile rhythms for haptic feedback, all dynamically adjusted based on IBS levels. Data is processed in real-time using MNE and HyPyP libraries, with post-study EEG analysis refining findings. The results aim to inform early-stage design guidelines for optimizing IBS through feedback mechanisms. Our key contributions are as follows:

\begin{enumerate}
  \item We determined the most effective feedback modalities and designs for enhancing IBS intensity, offering recommendations for designing such systems (Section \ref{result-summary}).
  \item We created an integrated system that measures IBS intensity in real-time using Emotiv EEG and Vicon motion capture for artifact removal, dynamically feeding the IBS values to visual, auditory, and haptic feedback systems (Section \ref{implementation}).
  \item We proposed the ``Cup \& String Task,'' a collaborative activity to test and analyze IBS under various feedback conditions (Section \ref{task}).
  \item We outlined practical design guidelines and explored potential applications in physical therapy and brainstorming activity (Section \ref{application}).
\end{enumerate}

\section{Background and Hypothesis}
Our work builds upon brain-to-brain interfaces, EEG-based hyperscanning in social interactions, and the exploration of different shared input modalities such as visual, audio, and haptic feedback.

\subsection{Inter-brain Sync and Joint Task}
Neuroscience literature has revealed that when two or more individuals interact, such as when watching a movie together~\cite{lankinen_intersubject_2014}, solving a puzzle jointly~\cite{fishburn_putting_2018}, playing an online game together~\cite{wikstrom_inter-brain_2022}, and even mimicking simple gestures between users~\cite{dumas_inter-brain_2010}, neural activity across their brains become synchronized. Recent studies reveal that synchronization rates vary significantly with the closeness of relationships between collaborators~\cite{kinreich_brain--brain_2017}, with most research focusing on shared goals~\cite{reinero_inter-brain_2021, fishburn_putting_2018, sanger_intra-_2012} rather than competitive settings~\cite{sinha_eeg_2016}. Synchronization is associated with various positive outcomes, including increased closeness, cooperation, prosocial behavior, and even improved team performance~\cite{wikstrom_inter-brain_2022}. 

%% Augmenting Inter-brain Synchrony
There has been research on wearable brain-to-brain systems aimed at augmenting IBS in natural settings, which identified user experiences such as hyper-awareness, relational interaction, and the dissolution of self~\cite{semertzidis_psinet_2024}. Our work, however, explores different input modalities, including shared Audio, Visual, and Haptic feedback.
\subsection{External Stimuli and Brainwave}
A number of works have explored the intrinsic impacts of external stimuli on brain wave activity ~\cite{nijholt_brain_2019, marian_cross-modal_2021}.
%For instance, studies discuss how intrinsic characteristics of visual, audio, and haptic stimuli affect brain activity, focusing on cross-modal interactions and their impact on memory retrieval~\cite{nijholt_brain_2019, marian_cross-modal_2021}. 

% Visual
Regarding \textbf{visual stimuli}, oscillatory visual patterns, like flickering lights or periodic changes ~\cite{spaak_local_2014}, or abstract visualizations ~\cite{sigrist_augmented_2013} have been demonstrated to be able to entrain brain rhythms ~\cite{de_graaf_alpha-band_2013, gomez-ramirez_oscillatory_2011, mathewson_making_2012}, offering insights into sensory processing and rhythmic control of attention and perception ~\cite{mathewson_making_2012, spaak_local_2014}. 
%Research indicates that abstract visualizations incorporating a limited set of applicable variables significantly enhance the comprehension of complex motor tasks~\cite{sigrist_augmented_2013}.
The IBS can also be impacted by visual stimuli, as previous studies demonstrated that even watching the same movie fosters viewers' IBS ~\cite{ahmadzadeh_does_2019}. Studies using fluctuating stimuli (e.g. dynamic and static visual stimuli) have shown that these can evoke distributed brain activities ~\cite{lu_influences_2016}.

% Auditory
On the \textbf{auditory side}, the relationship between sound and IBS has been explored through various activities, including speaking and listening~\cite{perez_brain--brain_2017} and musical activities~\cite{fujioka_beta_2009, gugnowska_endogenous_2022, cheng_brain_2024}. 
Other works investigate brain wave synchronization and modulation through periodic acoustic stimuli, including binaural beats and isochronic tones~\cite{will_brain_2007, aparecido-kanzler_effects_2021, ingendoh_binaural_2023}.

% Haptic
Prior research demonstrates that \textbf{haptic vibration} also influences various brain regions including the primary somatosensory cortex ~\cite{coll_cross-modal_2015, pisoni_cortical_2018, schirmer_touching_2019}, activating neural processes critical for sensory perception, memory, and attentional mechanisms ~\cite{alsuradi_eeg-based_2020}.
This has impacts on alpha, beta, theta bands, which are associated with sensory integration ~\cite{coll_cross-modal_2015, pisoni_cortical_2018, schirmer_touching_2019}, sensory accumulation, attentional selection ~\cite{chen_experimental_2017, park_neural_2019}, and memory load ~\cite{grunwald_theta_2001}.

% Summary
These studies collectively underscore the significant role of external stimuli in modulating brainwave activity. Inspired by these findings and implications, we implemented our feedback design for each modality that would foster changes in individual and IBS-related brain activities.
\\

\subsection{Interactive Feedback Systems}
\subsubsection{Visual Feedback}
Vision is a modality for spatial perception, supported by feedback strategies such as observation and imitation~\cite{sigrist_augmented_2013}. Visual feedback plays a clinical role in treating neurological disorders~\cite{ramachandran_use_2009}, and, when combined with physiological data such as EEG and heart rate, can enhance emotional regulation, mindfulness~\cite{dinh_fractalbrain_2024}, and relaxation~\cite{fernandez_deep_2019, amores_psychicvr_2016}.
Experimental approaches to visual cues expand mixed reality interactions ~\cite{kim_spatial_2024, kim_reality_2023}, using biosensory data to foster mindfulness, self-reflection, and relaxation in VR environments. Visualizing physiological data in VR has been shown to positively influence emotions and promote \textit{inter-brain synchronization}.

\subsubsection{Auditory Feedback}
Auditory feedback has been used to support motor learning~\cite{luciani_role_2022} and collaboration, such as jogging with shared spatial audio~\cite{mueller_jogging_2007}. Additionally, spatial audio improves shared experiences and assists in locating unseen objects~\cite{trepkowski_multisensory_2022}. Combining visual and audio feedback, Chen's research proposed an application for interpersonal synchrony in static scenarios~\cite{chen_hybrid_2021}.

\subsubsection{Haptic Feedback}
Interactive haptic devices play a crucial role in conveying sensations, physical skills, and subjective experiences such as emotion~\cite{bailenson_virtual_2007, hassib_emotion_2017} and presence~\cite{obrien_holding_2006, brave_intouch_1997, singhal_flex-n-feel_2017}. Examples include devices using vibration motors~\cite{lieberman_tikl_2007, miura_supporting_2006, shrestha_hdesigner_2023}, exoskeletons~\cite{heo_current_2012, maekawa_naviarm_2019, saraiji_fusion_2018}, mechanical linkages~\cite{nakagaki_linked-stick_2015, ivanova_lets_2017, ganesh_two_2014}, rollers~\cite{brave_intouch_1997}, ultrasonic methods~\cite{hutchison_non-contact_2008}, air pressure~\cite{suzuki_design_2002, suzuki_air_2005}, skin texture~\cite{kim_haptic_2018}, and electrical muscle stimulation (EMS)~\cite{duente_ems_2017, hanagata_paralogue_2018, nishida_biosync_2017, hassib_emotion_2017}.
Vibrotactile sensation is a particularly effective and accessible way to communicate haptic experiences over a distance, teaching motor skills such as rhythms~\cite{feng_augmented_2019}, melodies~\cite{huang_mobile_2010}, and gait movements~\cite{narazani_designing_2018}. Moreover, rhythmic vibrotactile stimuli can enhance cognitive functions, such as attention ~\cite{whitmore_improving_2024}.
%using wearables via haptic and multimodal rhythmic stimuli. 
This can also extend to sharing social aspects like social presence~\cite{heiss_enabled_2007} and heartbeats~\cite{heiss_enabled_2007, hassib_heartchat_2017, janssen_intimate_2010}, fostering togetherness over a distance~\cite{singhal_flex-n-feel_2017}. 
%Additionally, vibrotactile feedback, is a common interface already familiar to the public.
%, so it reduces potential confusion that may arise from novel haptic experiences.

\subsection{Team Performance and Communication}
%% Colloaboration Team Performance
Recent efforts to connect users more closely through technology have focused on building shared environments and notifying collaborators of changes~\cite{zhang_vrgit_2023, pejsa_room2room_2016}. Room2Room, for example, demonstrated effective meetings in virtual settings~\cite{pejsa_room2room_2016, gronbaek_blended_2024}. Studies also explore inter-brain coupling to enhance social presence in online games via biosignal sharing~\cite{hassan_augmenting_2024}, with VR collaborative tasks increasing brain synchrony~\cite{gumilar_comparative_2021}. Additionally, recent work emphasizes tailoring virtual environments to individual cognitive needs~\cite{kim_cognitive_2011}, while other research focuses on maintaining cognitive load to enhance teamwork awareness, highlighting the practicality of teamwork~\cite{antunes_developing_2011}. Furthermore, studies investigate the impact of virtual and real environments on collaboration~\cite{park_study_2019}, and ongoing efforts aim to enhance social presence by simulating users' physical characteristics and interaction styles~\cite{bourdot_social_2021}. Experimental findings also reveal differences in collaborative design performance and inter-brain synchrony between VR and real-world settings, highlighting how cognitive synchronization and communication influence teamwork effectiveness ~\cite{pan_exploring_2024}. Together, these elements expand the ways of communication in virtual interaction experiences~\cite{pai_neuraldrum_2020, lankinen_intersubject_2014, kim_dynamic_2023, kim_audience_2024}.

\subsection{Our Hypothesis and Approach}
To better understand the efficacy and design implications for feedback that can enhance IBS, we set the following hypothesis:
\begin{itemize}
    \item \textbf{H1}: Visual feedback increases IBS level
    %Visualization of IBS levels through on-body projection can enhance IBS %itself efficiently due to a low cognitive load
    \item \textbf{H2}: Auditory feedback increases IBS level
    %Audible frequencies-based feedback has an impact on IBS and can be used to enhance its levels
    \item \textbf{H3}: Haptic feedback increases IBS level
    %The rate of vibration as haptic feedback provides awareness of IBS levels and can be used to enhance IBS
    \item \textbf{H4}: External feedback can further augment IBS
    %Providing external feedback during collaboration can enhance IBS intensity between participants
\end{itemize}

% By bringing awareness to the current IBS state 
Motivated by prior works in IBS, feedback systems, and how external stimuli influence brainwaves, we aim to identify the most effective modality and feedback design for enhancing IBS in visual (H1), auditory (H2), and haptic (H3) modality, ultimately testing a hypothesis that providing external feedback during collaboration can enhance IBS intensity between participants (H4).

Within the same experiment condition, we controlled the frequency as its variable for every modality to sustain the fidelity of our comparison. Accordingly, our findings will provide early-stage design guidelines for optimizing IBS through feedback mechanisms.

Of the 5 common external senses, taste and smell are intrusive, have low temporal resolution, and are difficult to use as a feedback mechanism for this case. 
Additionally, visual, auditory, and haptic feedback are commonly used in many prior HCI research and commercial devices, and can easily be integrated into existing systems. Thus, we decided to test these modalities as a first step to understand their feasibility for IBS feedback.

\begin{figure*}[t]
    \centering
    \includegraphics[width=1.0\linewidth]{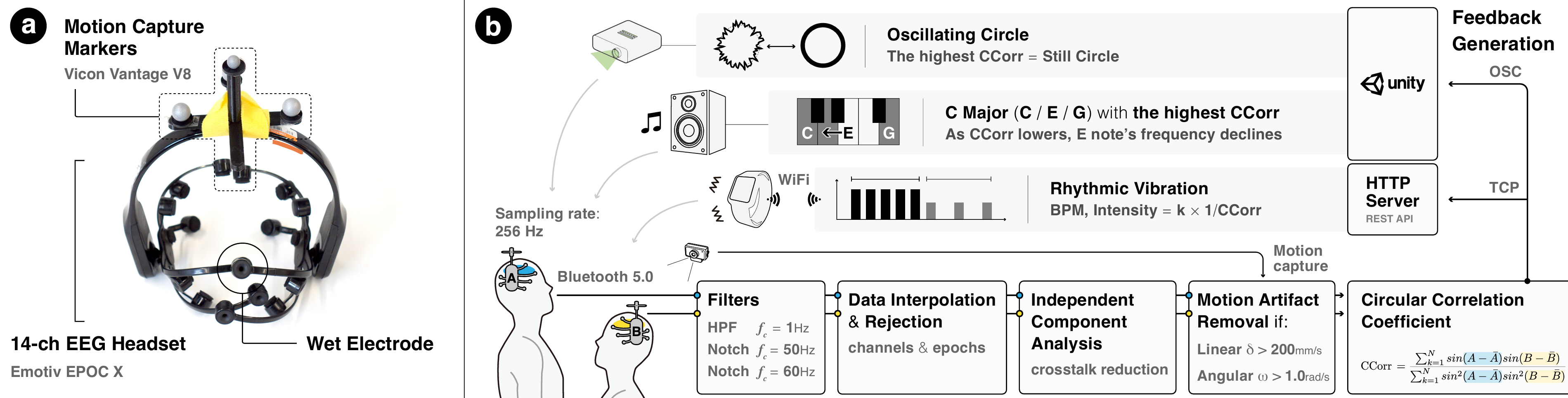}
    \caption{(a) Emotiv EPOC X EEG headset with mocap markers (b) Block diagram of the proposed feedback system}
    \Description{(a) Emotiv EPOC X EEG headset with mocap markers (b) Block diagram of the proposed feedback system}
    \label{fig:imp} 
\end{figure*}

\section{Implementation}
\label{implementation}
% sentence describing overall implementation
%Our work explores the integration of shared auditory, visual, and haptic feedback modalities into existing tasks. 
Our feedback system includes EEG measurement, a motion capture system, data analysis that encompasses preprocessing and IBS correlation estimation, and feedback systems. The details are provided below.

\subsection{Feedback Design}
The feedback system incorporates three types of external feedback—visual, auditory, and haptic—each designed to dynamically respond to IBS levels.

\subsubsection{Visual Feedback}
\paragraph{Hypothesis H1 Rationale}
Visual stimuli, such as oscillatory patterns ~\cite{de_graaf_alpha-band_2013, gomez-ramirez_oscillatory_2011, mathewson_making_2012}, flickering lights or periodic changes ~\cite{spaak_local_2014}, abstract visualizations ~\cite{sigrist_augmented_2013}, fluctuating stimuli ~\cite{lu_influences_2016}, shared media ~\cite{ahmadzadeh_does_2019}, has been shown to entrain brainwaves and offering insights into sensory processing and attention control ~\cite{mathewson_making_2012, spaak_local_2014}.

% Interaction rationale
On-body projection mapping is a technique where visual is presented on a user's body surface. This technique offers spatially aligned, low-cognitive-load visual feedback ~\cite{hoang_augmented_2017}, and often used for guiding physical movements ~\cite{amiri_keepstep_2021, xiao_mirrorfugue_2010, rogers_pino_2014, kirk_turn_2007}. Given its suitability for body motion feedback, we employed a simple oscillating circle as the on-body projection mapping for spatially congruent IBS level on both users' bodies (Figure \ref{fig:approach}a).

\paragraph{Feedback Design}
Our on-body projected orange-colored ring uses rhythmic visual oscillation patterns to respond to users' IBS metrics, transitioning between a smooth circle and a wave-like spiked surface (Figure \ref{fig:approach}a). Waves effectively represent continuous data, intuitively conveying changes over time as the ring dynamically projects visual feedback to capture and display the state of IBS (Implementation details in \ref{sec:implementation_visual_feedback}).

% Rationale: Sonification
\subsubsection{Auditory feedback} 
\paragraph{Hypothesis H2 Rationale}
Auditory stimuli can synchronize brainwave activity to specific frequencies ~\cite{ingendoh_binaural_2023, aparecido-kanzler_effects_2021, will_brain_2007}. Additionally, audio can convey spectral and temporal information without physical disruption~\cite{riecke_self-motion_2012}. It's also suitable for delivering dynamic feedback to multiple users in a shared space at the same time~\cite{hwang_effect-_2018}. 
% existing brain research has demonstrated that 
Moreover, the auditory cortex in our brain produces distinguishable responses to consonant and dissonant auditory stimuli created by simple frequency ratios~\cite{di_stefano_consonance_2022}.

As such, we incorporated a range of audible frequencies, including consonant and dissonant auditory stimuli. This allows users to perceive information without needing to visually focus on a specific location, simplifying the handling of multiple tasks and navigation in complex environments, while preserving the continuity of the experience.
%---
\paragraph{Feedback Design}
% Audio consonant and dissonant notes
In our approach, the IBS value is mapped to the frequency of a middle note in a musical chord (Figure \ref{fig:approach}b). A harmonic musical chord is used to guide the participants to perceive the shift of the middle note's frequency by comparing it with the other notes of the chord. Considering the rarity of individuals capable of discerning absolute frequency~\cite{deutsch_mystery_2019}, such design guides the participants to identify the IBS level by comparing the middle note to the first and third note of the chord using relative pitch detection as if the orchestral instruments' tunes their pitch before the concert (Implementation details in \ref{sec:implementation_audio_feedback}).
% Rationale: Vibration
\subsubsection{Vibrotactile Feedback}
\paragraph{Hypothesis H3 Rationale}
Haptic vibration influences various brain regions including the primary somatosensory cortex ~\cite{coll_cross-modal_2015, pisoni_cortical_2018, schirmer_touching_2019}, activating neural processes critical for sensory perception, memory, and attentional mechanisms ~\cite{alsuradi_eeg-based_2020}.
This has impacts on alpha, beta, theta bands, which are associated with sensory integration ~\cite{coll_cross-modal_2015, pisoni_cortical_2018, schirmer_touching_2019}, sensory accumulation, attentional selection ~\cite{chen_experimental_2017, park_neural_2019}, and memory load ~\cite{grunwald_theta_2001}.
These intrinsic responses underscore the ability of vibrotactile feedback to enhance perception, synchrony, and multisensory integration, leading to a higher IBS %through its provision.

% Interaction rationale
% Haptic feedback engages the sense of touch by providing a physical connection with their partners, including social presence ~\cite{heiss_enabled_2007, singhal_flex-n-feel_2017}, or conveying emotion ~\cite{heiss_enabled_2007, hassib_heartchat_2017, janssen_intimate_2010}.
Haptic feedback engages the sense of touch, providing a physical connection with partners that fosters social presence~\cite{heiss_enabled_2007, singhal_flex-n-feel_2017} and conveys emotion~\cite{heiss_enabled_2007, hassib_heartchat_2017, janssen_intimate_2010}. Vibrations are immediately felt (150 to 200ms ~\cite{luce_response_1991}), resulting in the user's attention being captured quickly. This is useful, especially in a situation where the visual resource is limited and user privacy should be prioritized ~\cite{pasquero_haptic_2011}.
%and a high association between the feedback and the user's experience of synchrony
%This has been used for conveying motor skills ~\cite{feng_augmented_2019, huang_mobile_2010, narazani_designing_2018},  from one to another user over distance.
% Feedback Selection
Thus, we employed rhythmic vibrotactile feedback due to the quick response times without diverting focus from the task, and is accessible to a broad community.

\paragraph{Feedback Design}
% heartbeat bpm - high = bad, low = good
We decided to provide vibration feedback that delivers rhythmic vibration with varying beats per minute (BPM) and intensities, reflecting levels of dissonance and synchrony, guided by prior work ~\cite{greem_variability_1974} on the haptic stimulus-sensation relationship (Figure \ref{fig:approach}c). A pilot study (N=5) informed the refinement of BPM and intensity intervals to enhance detectability and perceptual clarity during multitasking scenarios. (Implementation details in \ref{sec:implementation_haptic_feedback}).
%Furthermore, there has been research on wearable brain-to-brain systems by means of direct brain stimulation apparatus, aiming at augmenting the IBS ~\cite{semertzidis_psinet_2024}. 
%Our work, however, explores the possibility of using human fundamental modalities for enhancing the IBS, such as Visual (e.g. monitors or projectors), Auditory (e.g. speakers), and Haptic feedback (e.g. tactile wrist bands) that are easily incorporated into existing systems with widely available components and apparatus.

% Additionally, we investigated whether our system can further facilitate communication and the sharing of subjective feelings among multiple users in practical scenarios, such as performance art and team sports.

% Our aim is to guide the development of brain-to-brain interfaces toward a future that fosters human connection, while also examining how each input modality can impact and improve synchronization between users' brains. Our contributions are outlined below:
% In this paper, to determine which feedback modality has a significant effect on the IBS enhancement, 

\subsection{Dual-EEG data acquisition}
The EEG data was collected using two Emotiv EPOC X wireless EEG headsets, each equipped with 14 channels. The Emotiv EPOC X has been validated as a reliable, research-grade system for EEG studies~\cite{williams_its_2021}, and has been widely adopted in various brain activity analyses, including research on IBS~\cite{dikker_brain--brain_2017, chen_hybrid_2021} and other areas of neuroimaging~\cite{anderson_user_2011, williams_10_2020}. The headset's lightweight and wireless design makes it well-suited for the motor collaborative tasks we developed. From our pilot studies, we were able to observe the IBS phenomenon with Emotiv EPOC X and validate it through 6 different sets of headsets. Both headsets were connected to a single computer, with raw EEG data streams captured through an open-source Lab Streaming Layer (LSL)~\cite{kothe_lab_2024} protocol. All relevant data was collected across all 14 channels, using a sampling rate of 256 Hz.

\subsection{Real-time EEG Analysis}
We implemented a real-time EEG analysis to compute IBS circular correlation values, which were used to adjust the feedback. Our goal was to analyze participants' brain activity as they interacted with the system, utilizing two 14-channel raw EEG data streams. EEG data streams are fed into Python code utilizing the MNE library~\cite{gramfort_meg_2013} and the Hyperscanning Python Pipeline for inter-brain connectivity analysis (HyPyP)~\cite{ayrolles_hypyp_2020}. HyPyP provides various inter-brain connectivity measures with both automated and manual data preprocessing. This toolbox, widely used in IBS research, allows us to compute inter-brain connectivity~\cite{alimardani_robot-assisted_2022, schwartz_technologically-assisted_2022, turk_brains_2022, wikstrom_inter-brain_2022}, providing insights into neural mechanisms that underlie interactive experiences and how they facilitate communication and collaboration in mixed reality settings.

\subsubsection{Preprocessing}
\label{sec:preprocessing}

Preprocessing ensures that the EEG data is clean, interpretable, and reliable, improving the overall quality and accuracy of the subsequent analysis. In prior studies, the selection of epoch lengths of 1 to 1.5 seconds for inter-brain measure estimation is a common practice. This duration is often preferred as it provides a balance between accurately capturing the transient changes in neural activity and ensuring the dependability of the data obtained~\cite{dikker_brain--brain_2017, dumas_inter-brain_2010, reinero_inter-brain_2021}. However, recent studies have shown that shorter windows lead to unreliable estimations, particularly in cases of noisy signals~\cite{basti_looking_2022}. Real-time EEG raw data from two participants were recorded into two separate buffers of a sliding window of 3 seconds with a 1.5 seconds hop. The signals were then filtered with a band-pass filter for frequencies between 1 Hz and 48 Hz. The filtered data was used to calculate the IBS circular correlation value using HyPyP. In parallel, to ensure efficient real-time data preprocessing, we used Vicon motion capture to analyze the participants' head motion, as described in section \ref{sec:motion_artifact_removal}, to reject large and fast motions. While the system updates IBS every $\sim\!100\ ms$, the preprocessing algorithm is highly optimized with each process, with a total execution time of $\sim\!40\!-\!60\ ms$ for IBS circular correlation and $\sim\!2\ ms$ for Vicon based motion artifact classification (details in sec. \ref{sec:motion_artifact_removal}).

%Artifacts contamination in EEG signals is a common challenge in processing EEG data. Some common artifacts include motion, ocular, muscular, and cardiac ~\cite{bajaj_wavelets_2020}. The HyPyP toolbox includes preprocessing, including adjustable filters, slow drift removal, rejecting and interpolating the inadequate channels and highly noisy epochs, and independent component analysis. 
%
%sampling at 256 Hz, freq: [1, 48] (0.16 – 43Hz, notch filters at 50Hz and 60Hz - innate) FFT size, Window size, Hop size

\subsubsection{Motion Artifact Removal}
\label{sec:motion_artifact_removal}
\begin{figure}[h]
    \centering
    \includegraphics[width=1.0\linewidth]{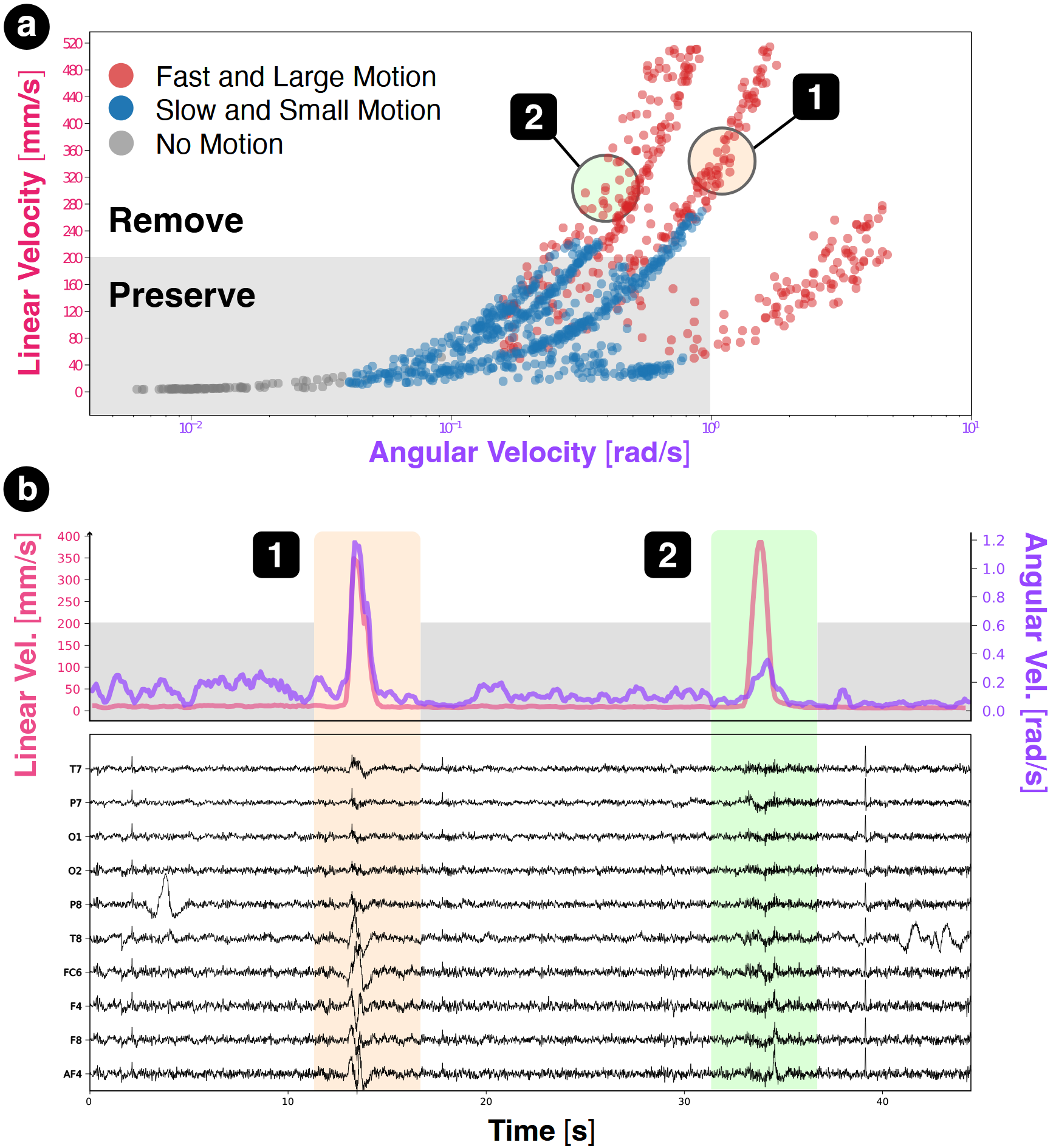}
    % \caption{Characterization of motion artifacts while people wearing the EEG headset staying still as our baseline that we call \textit{No Motion} (gray), performing \textit{Slow and Small Motion} (blue), and \textit{Large and Fast Motion} (red). Left side plot (a) shows motion plot linear velocity (magenta) vs. angular velocity (purple). The values for which motion is rejected (removed) or preserved are segmented with a gray box. The right side plot (b) shows in the upper section, linear velocity (magenta) and angular velocity (purple) over time and in the bottom section, EEG channel response with motion artifacts for (1) head tilt example of a large and fast angular motion highlighted in orange and (2) body shifting example of linear motions highlighted in green.}
    %\caption{Characterization of motion artifacts in EEG recordings under three conditions: (a) No Motion (gray), Slow and Small Motion (blue), and Large and Fast Motion (red). (b) Top: Motion over time with rejected segments are shown linear motion (magenta) and angular velocity (purple). Bottom: EEG channel response with motion artifacts, highlighting large/fast motion (1) head tilt (orange, angular motion) and (2) body shifting (green, linear motion).}
    \caption{(a) Motion artifacts in EEG recordings: No Motion (gray), Slow/Small Motion (blue), and Large/Fast Motion (red). 
    (b) Top: Motion over time with rejected segments highlighted: linear motion (green) and angular velocity (orange). Bottom: EEG channel response with motion artifacts.}
    \Description{(a) Motion artifacts in EEG recordings: No Motion (gray), Slow/Small Motion (blue), and Large/Fast Motion (red). (b) Top: Motion over time with rejected segments highlighted: linear motion (green) and angular velocity (orange). Bottom: EEG channel response with motion artifacts.}
    \label{fig:MoArtRemove}
\end{figure}

Considering the gestural nature of the collaborative tasks, we also augmented the EEG headset with the Vicon motion capture (MoCap) markers, as shown in Fig. \ref{fig:imp}b, to track both linear and angular fast and large motion and reject them from our calculations, see Fig. \ref{fig:MoArtRemove}b. The Vicon MoCap system is submillimeter accurate~\cite{merriaux_study_2017}, sampled at 100 Hz, and gives both linear and angular position. To characterize undesired motion, we conducted a pilot study with participants moving with slow and fast linear and rotational motions along with EEG data. We segment EEG artifacts and their corresponding motions to establish classifier motion parameters of over 200mm/s linear velocity and 1 rad/s angular velocity as shown in Fig. \ref{fig:MoArtRemove}. Both participants' time segments were evaluated, and if a segment was rejected due to motion artifacts, the system reverted to outputting the CCorr value from the previous epoch.

\subsubsection{Measuring Inter-brain Synchronization}
After pre-processing, the system calculated the level of IBS between two participants. Although Phase Lock Value (PLV) and Phase Lock Index (PLI) were widely used in previous research to measure IBS~\cite{hsu_cooperative_2021, dumas_inter-brain_2010, schwartz_generation_2024, perez_brain--brain_2017}, however, Burgess et al.~\cite{burgess_interpretation_2013} suggested that CCorr shows more robustness to coincidental synchrony. The CCorr measures IBS by assessing the alignment of phase angles in brain signals between individuals, capturing how neural oscillations are synchronized during shared activities. High covariance and CCorr values near 1 indicate strong synchrony, while low covariance and CCorr values approaching 0 suggest minimal synchrony~\cite{honari_evaluating_2021}.

CCorr for two corresponding EEG channels is: 
% equation of ccorr
\begin{equation}
    \mathrm{CCorr}_{A,B} = \frac{\sum_{k=1}^{N}sin(A-\bar A)sin(B-\bar B)}{\sum_{k=1}^{N}sin^2(A-\bar A)sin^2(B-\bar B)}
\end{equation}

where $A$ and $B$ are the phases of two EEG signals, and $\bar A$ and $\bar B$ are the average phases of respective signals. 

For each buffer of 3 seconds, the system calculated channel-to-channel CCorr from two participants. From the 14 CCorr values derived from each channel, the average of the five highest CCorr values is used as the IBS metric. To mitigate bias introduced by averaging multiple correlation coefficients, the top 5 pairwise CCorr values were transformed using Fisher's z-transformation~\cite{semertzidis_psinet_2024}. The z-transformed values were averaged, and this average was subsequently converted back using the inverse Fisher's z function. Fisher's z-transformation is defined as follows:

\begin{equation}
    z = \frac{1}{2} \log_e\left(\frac{1+r}{1-r}\right)
\end{equation}

With the buffers of 3 seconds sliding window with a 50\% overlap~\cite{dikker_brain--brain_2017, dumas_inter-brain_2010, reinero_inter-brain_2021}, the system outputs a CCorr value every 1.5 seconds. Based on our pilot study of 16 sessions, we observed that the IBS metric was approximately 14\% higher during collaborative tasks compared to non-collaborative situations.

% derivation ~\cite{ayrolles_hypyp_2020}
% referable ~\cite{wikstrom_inter-brain_2022}

\subsection{Feedback System}
Through the proposed IBS metric, the average of the top five highly correlated CCorr values from 14 channels are updated every 1.5 seconds. The metric is sent via OSC~\cite{wright_open_2005} and WIFI to each feedback platform. Three different types of modalities, audio, visual, and haptic, are designed as the feedback. The level of feedback was divided into five subdivisions according to the IBS metric.  

\subsubsection{Visual Feedback: Projection Mapping}
\label{sec:implementation_visual_feedback}

\begin{figure}[h]
    \centering
    \includegraphics[width=1.0\linewidth]{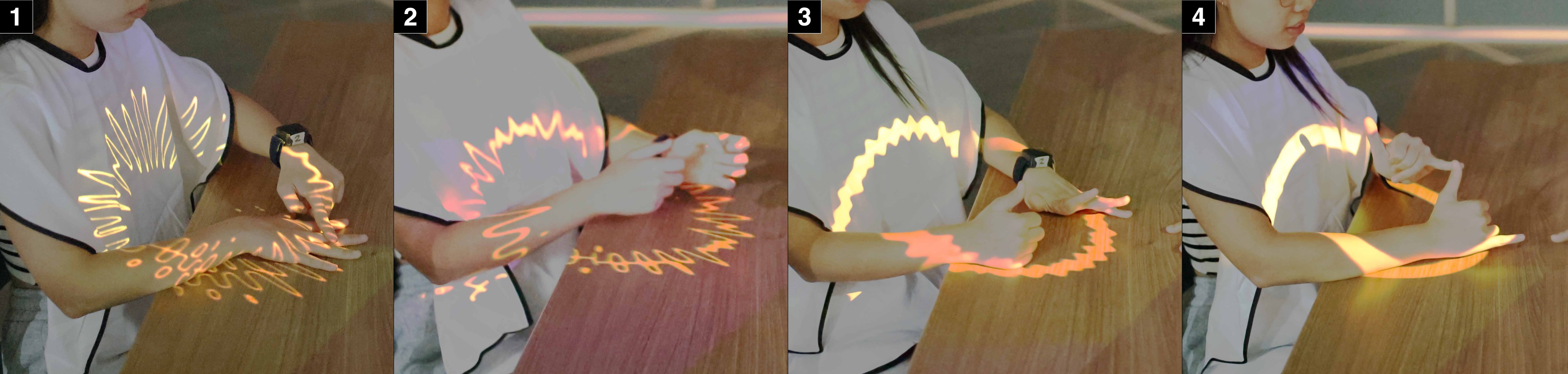}
    \caption{Different levels of IBS are illustrated by a ring-shaped visual representation. The amplitude of the waveform corresponds to the IBS intensity, where the sync levels are: (1) the lowest, (2) low, (3) high, and (4) the highest.}
    \Description{Different levels of IBS are illustrated by a ring-shaped visual representation. The amplitude of the waveform corresponds to the IBS intensity, where the sync levels are: (1) the lowest, (2) low, (3) high, and (4) the highest.}
    \label{fig:vis}
\end{figure}

%Research indicates that abstract visualizations incorporating a limited set of applicable variables significantly enhance the comprehension of complex motor tasks~\cite{sigrist_augmented_2013}. 
The proposed application creates an abstract visualization of an oscillating circle using Unity that is controlled by the IBS value. The oscillation has a negative relationship with the IBS value; as the IBS value increases, the oscillation decreases (Fig. ~\ref{fig:vis}-1). Consequently, a perfect IBS value would result in a still circle (Fig.~\ref{fig:vis}-4), while a low IBS value would correspond to increase oscillation. To project the visual, we used a pair of LG PF50KA portable projectors with 600 ANSI lumens brightness. Our on-body projection mapping approach makes the system expandable to the multiuser feedback in a further group collaboration scenario. 

\subsubsection{Auditory Feedback: Frequency Shift}
\label{sec:implementation_audio_feedback}
%The existing brain research has shown that consonant and dissonant auditory stimuli caused by simple frequency ratios cause distinguishable responses in the auditory cortex of our brain~\cite{di_stefano_consonance_2022}. Our Unity application creates an audio controlled by the IBS metric. The IBS value is mapped to the frequency of a note that determines the amount of consonance and dissonance. A harmonic musical chord is used to guide the participants to perceive the shift of the mid-notes frequency perceivably. A chord is a group of three or more notes played simultaneously. Especially the harmonic series chord is a sequence of frequencies that are integer multiples of a fundamental. The utilized Major triad chord is composed of three notes with a 4:5:6 frequency ratio. Among the three notes, the first (root) note and the third (perfect fifth) note have a 2:3 ratio, which is a simple, low-integer ratio that creates a highly stable and consonant sound. Our auditory system processes sounds in ways that make certain intervals feel more "stable" or "pleasant." 

In our approach, the IBS value is mapped to the frequency of a middle note in a musical chord. A harmonic musical chord, a group of three or more notes played simultaneously, is used to guide the participants to perceive the shift of the mid-notes frequency perceivably. Especially, the harmonic series chord is a sequence of frequencies that are integer multiples of a fundamental. The utilized Major triad chord has three notes with a 4:5:6 frequency ratio. Among the three notes, the first (root) note and the third (perfect fifth) note have a 2:3 ratio, which is a simple, low-integer ratio that creates a highly stable and consonant sound. Our auditory system processes sound in ways that make certain intervals feel more ``stable'' or ``pleasant~\cite{wright_mathematics_2009}.''

To induce this effect, the middle note’s frequency shifts from the major third to a lower frequency based on the IBS level. One level degradation of IBS corresponds to a 5\% decline of the middle note’s frequency distorting the ratio of harmonic series more noticeably. While the just-noticeable difference (JND) is about 3 -- 5 Hz around 500 Hz -- 600 Hz, our frequency shift occurs between 547 Hz -- 659 Hz over 5 different levels~\cite{houtsma_pitch_1990}. 
During the feedback, our Unity code synthesizes the corresponding notes as a chord and plays them every 1.5 seconds, synchronized with the IBS metric update. The audio is created by the 30 watt stereo speakers in front of the participants.

\subsubsection{Haptic Feedback: Vibration Band}
\label{sec:implementation_haptic_feedback}

We utilized a haptic band as shown in Fig. ~\ref{fig:rep}d that builds upon~\cite{shrestha_hdesigner_2023} to provide vibrotactile feedback in the form of repeating beats where faster rate in the form of beats per minute (BPM) and stronger intensities indicating higher dissonance while slower BPM and lower intensities indicating higher synchrony. The feedback was divided into five subdivisions with increment following~\cite{greem_variability_1974} characterization of the haptic stimulus-sensation relationship.

We conducted a pilot study (N=5) in which participants provided a thumbs-up or thumbs-down signal to indicate whether they perceived the vibrations as faster and more intense or slower and less intense, respectively. Participants were instructed to disregard instances where they were unsure of their perception. We systematically varied BPM and intensity values and observed participants' ability to discern changes. Initially, we tested 20 intervals, informed by~\cite{greem_variability_1974}, but refined this to 10 intervals based on experimental findings. To ensure consistency across modalities and enhance perceptual distinction, we decided the intervals to five in the final study.

The haptic band (hBand v2~\cite{shrestha_hdesigner_2023}) hardware is manufactured by vAIolin~\cite{vaiolin_software_2025}. It is powered by an ESP8266 microcontroller with WiFi connectivity pulse-width modulation (PWM) output connected to four eccentric rotating mass (ERM) vibration motors (VC1030B028F). The device is compactly integrated into a wearable watch-like 3D printed form factor weighing 200 g, powered by a LiPO battery. 
%Once the device is turned on, it connects to the paired PC via Wifi connectivity. 
The devices accept REST API to control the vibration patterns as described in ~\cite{shrestha_hdesigner_2023}.

\subsubsection{Post-study EEG Analysis} \label{post-study analysis}
To assess the impacts of our interactive media system on human interaction and expression, we require multichannel EEG analysis for both real-time feedback and post-analysis. Raw EEG data were extracted from each participant's recordings. After applying a band-pass filter between 1 Hz and 48 Hz, the continuous EEG data were segmented into 3-second epochs with a hop size of 0.5 seconds. To minimize the influence of participants re-adjusting themselves causing noisy EEG data at the start and end of each trial, we excluded the first and last 3 epochs from each task condition.

% Filtering musclar activity using spectral energy density
In order to address further artifact contamination, we first transform the EEG time-domain signal to a frequency domain signal by performing a short-time Fourier transform (STFT) and compute the \textit{spectral centroid} to measure the energy density of a discrete-time window for each channel~\cite{klapuri_signal_2006}. The cumulative sum of all channels is used to filter out epochs with high energy indicative of noise beyond the neural activity usually as a result of muscular activity such as eye movements, jaw clenching, and eyebrow movements. We rejected any trial with less than 50\% valid epoch~\cite{dikker_brain--brain_2017}.

After preprocessing and rejecting all invalid trials, the remaining epochs were used to calculate the average IBS CCorr through Fisher's z-transformation. This approach mirrors the real-time CCorr calculation method described in the previous section.
% which computes the discrete Fourier transform (DFT) over short sliding window of 256 samples and hop length of 32 \cite{librosa} on each channel. - don't have to cite librosa. everyone uses it

% Hypothesis 1 (H1): To validate our control condition, we perform paired t-test between the sync and non-sync no-feedback condition trials. 
% Hypothesis 2 (H2): Feedback forsters IBS (Ha) Audio, (Hb) Visual, and (Hc) Haptics
% Hypothesis 3 (H3): Pairs with same preference for feedback modality, will synchronize better than pairs with different feedback preferences.
% FUTURE WORK
%a) individual preference single feedback to each pair 
%b) Multi-modal but same feedback to both pairs.

\section{User Study}
%\subsection{System Validation}
%To validate whether our hardware and software configuration can detect the IBS between two participants and to explore the effect and role of each feedback modality on IBS enhancement, we set the following five conditions: Non-sync, No Feedback, Visual Feedback, Auditory feedback, and Haptic Feedback conditions. We then evaluated the IBS value for each condition. Our study was approved by the Institutional Review Board (anonymous number).

\begin{figure*}[h]
    \centering
    \includegraphics[width=1.0\linewidth]{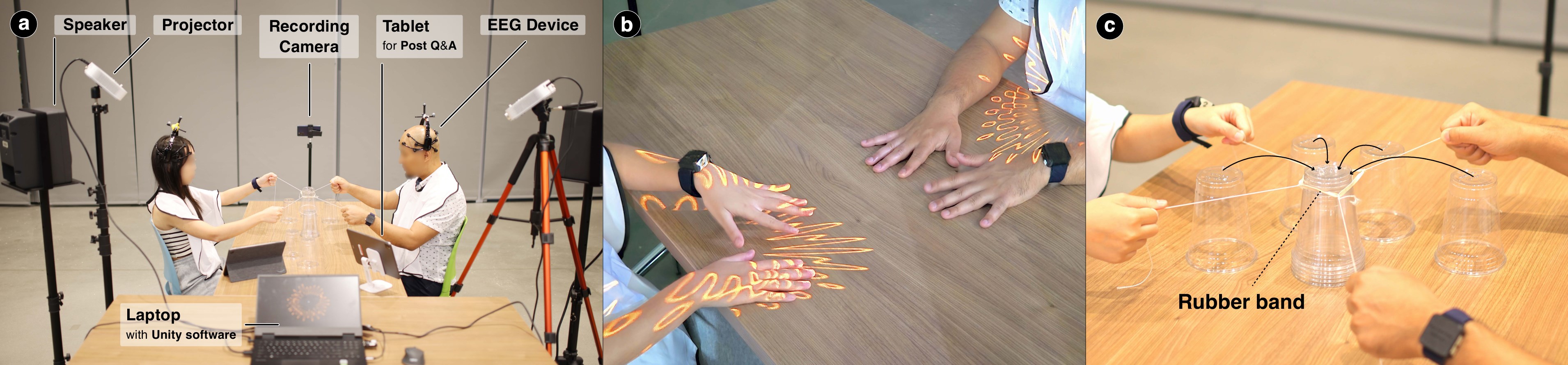}
    \caption{(a) User study setup (b) Hand Mimicry session (c) Cup \& String session}
    \Description{(a) User study setup (b) Hand Mimicry session (c) Cup \& string session}
    \label{fig:US_setup}
\end{figure*}

\subsection{Tasks}
\label{task}
Participants were asked to perform joint action tasks that consisted of a Hand Mimicry session and a Cup \& String session. The Hand Mimicry was chosen based on the study from Dumas~\cite{dumas_inter-brain_2010} that shows the spontaneous imitation of hand movements improves the IBS (Fig. ~\ref{fig:US_setup}b). Similarly, during the Cup \& String session, participant pairs were asked to manipulate the cups only using the strings and a rubber band, and the task was considered complete when all cups were stacked or unstacked in the desired order (Fig. ~\ref{fig:US_setup}c). These challenges were chosen considering the task's cooperation, coordination, joint attention, and participants' body condition~\cite{valencia_what_2020}.

\subsection{Conditions}
There were a total of 10 trials, which involved 2 joint action tasks combined with 5 conditions: Non-sync, No Feedback, Visual Feedback, Auditory Feedback, and Haptic Feedback.

\begin{itemize}
    \item In the \textbf{Non-sync} condition, participants were instructed to watch a video with significant personal meaning, designed to immerse them fully. This approach ensures variability in their attention, physical state, and task engagement. 
    \item In the \textbf{Sync} condition, participants engaged in two joint action tasks as described previously. Each task was performed under four distinct feedback conditions: No Feedback, Visual Feedback, Auditory Feedback, and Haptic Feedback. 
    \begin{itemize}
        \item In the \textbf{No Feedback} condition, participants completed the task without receiving any feedback on their IBS levels. 
        \item Conversely, in the \textbf{Visual}, \textbf{Auditory}, and \textbf{Haptic} conditions, participants received real-time feedback on their IBS levels through visual, auditory, and haptic cues, respectively. 
    \end{itemize}
\end{itemize}

We measured the CCorr value for each trial. The two Non-sync trials served as system validation, and were implemented to ensure (a) the reliability and interpretability of our EEG recordings, and (b) the effectiveness of our circular correlation calculations in capturing IBS. Based on previous research regarding inter-brain synchrony for cooperation and collaboration~\cite{dikker_brain--brain_2017, dumas_inter-brain_2010}, we anticipated that CCorr values would be higher in the Sync condition compared to the Non-sync condition. All combinations of tasks and conditions were performed in a randomized order in each study session. Each task lasted for 1-2 minutes, following a rest and setup time of 2-3 minutes between consecutive tasks. Before each trial, participants will be given time to practice with the feedback modality to familiarize themselves with its interpretations. Our study was approved by the Institutional Review Board (anonymous number).

\subsection{Setup}
In the experiment room, participants were seated facing each other across a rectangular table. Both participants were fitted with a 14-channel wireless EMOTIV EPOC X EEG headset. The signal quality and EEG data quality were determined based on inspection using EMOTIV's proprietary software. Participants wore the EEG device throughout the whole study. After the EEG setup, we connected two EEG headsets to a main computer in the experiment room and acquired the raw EEG stream using the LSL. Two experimenters facilitated and oversaw the experiments. All sessions were recorded on video and audio as well.

\subsection{Questionnaires}
Questionnaires are composed of the inter-trial questionnaire and post-study questionnaire.
\subsubsection{Inter-trial Questionnaire}
The questionnaire included selected relevant questions from the Cross-Jurisdictional Sharing (CJS) Agreements Collaborative Trust Scale~\cite{bullard_improving_2008} to better fit our study, as well as additional questions to evaluate the type of feedback. We used a modified version of the CJS Agreements Collaborative Trust Scale. Our trust scale score included: \textit{The collaborative partners are motivated to protect our common interests}; \textit{The collaborative partners share a common vision of the end goal of what working together should accomplish}; \textit{I found this type of feedback to be helpful with synchronizing with my teammate}; \textit{I found this type of feedback easy to understand for IBS}.

\subsubsection{Post-study Questionnaire}
At the end of the study session, participants were asked to make ratings of the three types of feedback on a 7-point Likert scale that went from 1 = The Worst, 4 = Neutral, and 7 = The Best. This will be followed by an open-ended interview to gather more detailed feedback. We explored participants' reasoning behind their ratings, how the feedback influenced collaboration between the pair, and any personal factors that may have shaped their preferences for each type of feedback. 

\subsection{Participants}
We recruited 34 participants (20 female, 14 male; Mean 23 $\pm$ 4 years old), organized into 17 pairs. Participants were recruited via university email lists and word of mouth, and were asked to complete a screening survey to ensure eligibility. Participants had to be adults over 18, with normal or corrected vision, no neurological or psychiatric history, and able to attend the study in person. Individuals with visual, auditory, or motor impairments were excluded due to the nature of the feedback and motor tasks.

Eligible participants were randomly paired based on scheduling availability, with 41\% of the pairs being mixed-gender and 59\% same-gender. Out of the 17 pairs, one pair knows each other, while the remaining 16 pairs are unfamiliar with each other. Each session lasted no longer than 90 minutes, including setup time, and participants received \$20 as compensation. Written informed consent was obtained from each participant prior to the experiment. One pair was excluded from post-study analyses due to excessive EEG noise. 

\section{Results}
We conducted a quantitative analysis based on the IBS CCorr values by the condition and qualitative analysis from the participants' interviews.
\subsection{Quantitative Results}
To evaluate the impact of these feedback modalities, we performed several statistical analyses, including normality checks, paired t-tests, and non-parametric tests, with appropriate p-value corrections. We conducted a study with 17 pairs of participants who performed tasks under four experimental feedback conditions (No Feedback, Auditory, Visual, and Haptics), along with a Non-sync condition. The results are from the~\textit{Cup \& String} and \textit{Hand Mimicry} sections of the tasks to evaluate the overall effects of feedback modalities. We performed several statistical analyses, including normality checks, paired t-tests, non-parametric tests, and p-value corrections using the Benjamini-Hochberg (BH) correction.

% We calculated the Average IBS circular correlation values for each pair in each condition using method mentioned in \ref{post-study analysis}.

\begin{figure}[h] 
    \includegraphics[width=1.0\linewidth]{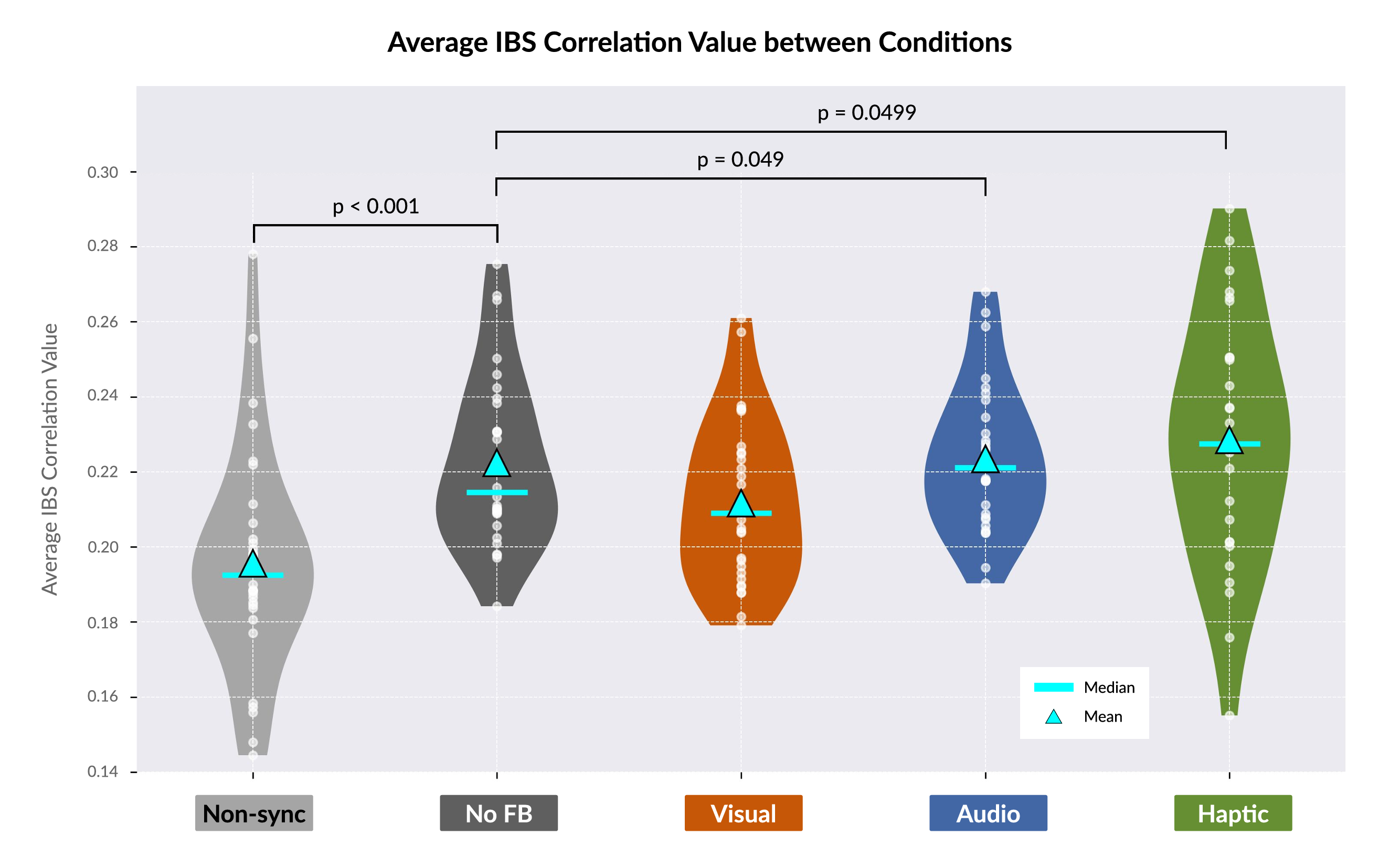}
    \caption{Average IBS Circular Correlation Value with three feedback modalities across all tasks. Each dot represents a CCorr value within that condition. Results showed significant differences Non-sync and No Feedback (No FB) validating our setup of shared task substantially enhancing IBS. The comparison between No Feedback with Auditory was statistically significant and Haptics was marginally significant.}
    \Description{Average IBS Circular Correlation Value with three feedback modalities across all tasks. Each dot represents a CCorr value within that condition. Results showed significant differences between Non-sync and No Feedback (No FB) validating our setup of shared tasks substantially enhancing IBS. The comparison between No Feedback with Auditory was statistically significant and Haptics was marginally significant.}
    \label{fig:Average IBS}
\end{figure}

\subsubsection{Non-sync vs. Sync}
As shown in Fig.~\ref{fig:Average IBS}, the Sync condition in our setup produced significantly higher mean and median IBS CCorr values compared to the Non-sync condition (SD = 0.0292, p = 0.0068). These results align with prior research, confirming that our hardware and software configuration can detect the IBS between two participants, and our shared task setup effectively enhances inter-brain synchrony. This also aligns with our findings of a trend in our trust score~\cite{corbin_basics_2015}. In our analysis, a Chi-square test was conducted to compare the feedback modalities (Visual, Audio, Haptic) against the No feedback mode. The test yielded a Chi-square statistic of approximately 108.75 with 12 degrees of freedom (df = 12) and a sample size of n = 680. The p-value was 0.0643, suggesting a marginal increase with the feedback modality. The effect size, measured by Cramér’s V, was 0.2308, indicating a small to moderate association between feedback modalities and trust scores~\cite{mchugh_chi-square_2013}, suggesting the association between IBS and trust scores.

\subsubsection{Normality Check}
Before conducting the statistical tests, we performed a Shapiro-Wilk normality test for each condition to assess whether the data followed a normal distribution. The p-values for the Non-sync (p = 0.068), No Feedback (p = 0.083), Auditory (p = 0.218), Visual (p = 0.265), and Haptics (p = 0.961) conditions all exceeded the 0.05 significance threshold, indicating that the data in each condition are normally distributed. Given these results, parametric tests such as paired t-tests were appropriate for the subsequent analysis.

\subsubsection{Parametric Comparison to Non-sync}
Paired t-tests were conducted to compare each feedback condition (No Feedback, Auditory, Visual, and Haptics) against the Non-sync. The results showed significant differences for all feedback conditions. To control for the risk of Type I errors due to multiple comparisons, we applied the Benjamini-Hochberg (BH) correction with a False Discovery Rate (FDR) of 0.10~\cite{perneger_whats_1998, benjamini_controlling_1995} and all comparisons with the Non-sync remained statistically significant, specifically, No Feedback (t(32) = -5.739, p < 0.001), Auditory (t(32) = -2.971, p = 0.0058), Visual (t(32) = -5.347, p < 0.001), and Haptics (t(32) = -7.345, p < 0.001). These results indicate that all feedback conditions had a significant impact compared to the Non-sync, even after adjusting for multiple comparisons.

\subsubsection{Parametric Comparison to No Feedback}
To explore differences between the feedback modalities, paired t-tests were conducted comparing Auditory, Visual, and Haptics directly against No Feedback. The results indicated that only the Auditory condition was significantly different. Visual and Haptics were not significantly different from No Feedback. After applying the BH correction FDR of 0.10~\cite{perneger_whats_1998, benjamini_controlling_1995}, the comparison between No Feedback with Auditory remained significant (t(32) = 2.555, p = 0.0049), Haptics became marginally significant (t(32) = -0.986, p = 0.0499). However, Visual (t(32) = 0.118, p = 0.0907) was not significant after correction. Thus, the Auditory and Haptics conditions showed a significant difference from No Feedback after p-value correction, while Visual did not demonstrate significant differences.

\subsubsection{Non-parametric Comparison to No Feedback}
To validate the results and account for any deviations from normality, we conducted a Wilcoxon signed-rank test as a non-parametric alternative. The test yielded consistent results with the paired t-tests, confirming that the Auditory feedback condition remained significantly different from No Feedback, while Visual and Haptics did not show significant differences. After applying the BH correction with FDR of 0.10~\cite{perneger_whats_1998, benjamini_controlling_1995} to account for multiple comparisons, the Auditory condition remained significant (t(32) = 109.5, p = 0.0057), Haptic became significant (t(32) = 165.5, p = 0.0398), while Visual (t(32) = 196.5, p = 0.0902) remained non-significant. These findings confirm the robustness of the analysis, indicating that the observed differences are not dependent on the assumption of normality or multiple comparison issues.

\subsubsection{Effect Size}
To further explore the practical significance of the differences, we calculated effect sizes using Pearson’s correlation coefficient for the comparisons between No Feedback and the other modalities. After applying the BH correction with FDR of 0.10~\cite{perneger_whats_1998, benjamini_controlling_1995} to control for multiple comparisons, the analysis revealed a small effect for Auditory feedback (r = 0.184, corrected p = 0.006), indicating that Auditory feedback had a slight but significant impact on performance compared to No Feedback. In contrast, the effect sizes for Visual feedback (r = 0.0005, corrected p = 0.907) and Haptics feedback (r = 0.032, corrected p = 0.050) were negligible, suggesting that these feedback modalities had minimal to no meaningful effect on performance when compared to No Feedback.

% The effect size between No Feedback and Auditory was moderate (r = 21.07), suggesting a notable, albeit non-significant, difference after correction. The effect sizes for Visual and Haptics were smaller (r = 37.82 and r = 31.85, respectively), indicating minimal differences from No Feedback. This suggests that Auditory feedback had the most notable impact, while the effects of Visual and Haptics feedback were less substantial.

\subsubsection{Summary} 
\label{result-summary}
The findings suggest that Auditory feedback had the most consistent and meaningful effect compared to No Feedback, both statistically and in terms of effect size. In contrast, Visual and Haptics Feedback did not show significant improvements over No Feedback, even though their effect sizes were relatively large. Overall, the Auditory feedback emerged as the most impactful feedback modality in this study.
% \removed{These results allow us to test our hypotheses, as shown in Table~\ref{Tab:hypothesis}. Specifically, (H1) Visual feedback did not foster the IBS metric, while (H2) Auditory and (H3) Haptic feedback showed a positive influence on the IBS metric. The feedback in general (H4) also further augmented the IBS.}

\subsection{Qualitative Results}
Through the utilization of grounded theory (GT) methodologies, we examined the open-ended responses collected from interviews~\cite{corbin_basics_2015}. This qualitative research strategy curbed bias and revealed patterns, motives, and justifications behind participants' sentiments.
%, encapsulating a vast spectrum of thoughts and not simply focusing on recurring themes.} 
%
To identify trust-related trends, our study explored individual experience aspects like trust, intimacy, and empathy. Unexpectedly, we discovered new concepts during the review. For example, one participant noted that incorrect feedback mechanisms often gave a false sense of synchronization, impacting their experience.
We categorized this new concept and examined its occurrence across various feedback mediums. This approach helped us identify false synchronization as a crucial factor influencing user perception.
%Our method allowed us to derive insights from real participant experiences, ensuring our conclusions were rooted in reality.}

\begin{figure}
    \centering
    \includegraphics[width=1.0\columnwidth]{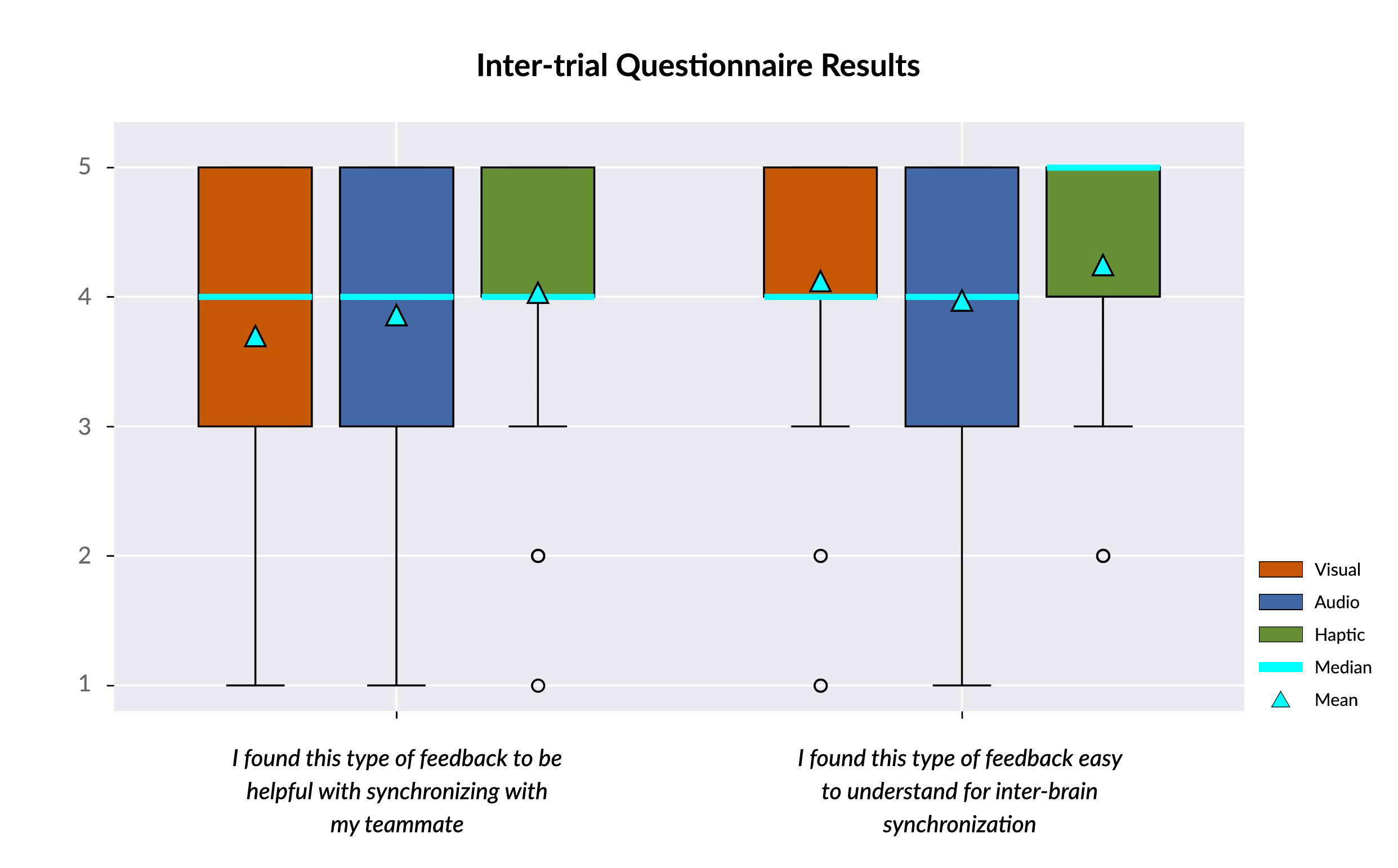}
    \caption{Inter-trial Questionnaire Results: (Left) I found this type of feedback to be helpful with synchronizing with my teammate (Right) I found this type of feedback easy to understand for inter-brain synchronization}
    \Description{Inter-trial Questionnaire Results: (Left) I found this type of feedback to be helpful with synchronizing with my teammate (Right) I found this type of feedback easy to understand for inter-brain synchronization}
    \label{fig:Likert}
\end{figure}

\subsubsection{User Preference on Sensory Reaction}

% \begin{table}[h]
% \centering
% \captionof{table}{Participants' quote from user study}
% \begin{tabular}{|p{1.5cm}|p{1.5cm}|p{8cm}|}
% % \begin{tabular}{|c|c|p{8cm}|}
% \hline
%     \textbf{Feedback} & \textbf{Participant} & \textbf{Quote}             \\ \Xhline{1.5pt}
%     Audio \centering & P23\centering\arraybackslash & ''Audio feedback was nice because it provided instant, delay-free responses. Performing tasks with audio in the background worked really well, allowing full focus on the task while receiving feedback.'' \\ \hline
%     Visual\centering\arraybackslash & P04\centering\arraybackslash & ''When I really focus on the task, the Visual feedback fades into the background, and I stop seeing them.''   \\ \hline
%     Haptic\centering\arraybackslash & P07\centering\arraybackslash & ''Initially haptic felt nice and felt more sync, but later on I was paying too much attention to haptic so it could have been distracting.''   \\ \hline
% \end{tabular}
% \label{Tab:hypothesis}
% \end{table}

In analyzing the reaction from participants to sensory feedback mechanisms — visual, audio, and haptic — three major themes emerged.
%The study aimed to evaluate the effectiveness of these feedback types in facilitating synchronization during collaborative tasks and to understand their influence on user experience and task performance. By delving into the participants' experiences and preferences, the study sought to uncover insights that could inform the design of more effective feedback systems for various collaborative contexts.
The results revealed that preferences were evenly distributed and showed strong polarized opinions: Haptic feedback was both the most and least favored, with 13 people rating it as their favorite and 14 as their least favorite. Visual feedback also exhibited polarization, with 12 people favoring it and 11 people considering it their least favorite. Audio feedback was the least polarizing method with 9 people rating it as their favorite and 9 as their least favorite. The following sections investigate these results further by articulating three themes: polarized opinions on Haptic feedback, polarized preferences for Visual feedback, and balanced acceptance of Audio feedback.

%Feedback preferences varied based on personal and professional backgrounds, with musicians favoring Audio feedback and those with visual-spatial skills preferring Visual feedback. The nature of the task also mattered; rapid synchronization tasks benefited from Audio feedback, while simpler low-stakes tasks suited Visual feedback. Haptic feedback might be more suitable for stressful tasks, as mentioned by participants. This necessitates customizable feedback systems.

\textbf{Audio feedback} presented both advantages and challenges. It was generally seen as distinct and beneficial, particularly for tasks requiring rapid synchronization, as its immediacy enabled quick adjustments. However, some participants found it intrusive or confusing if not perfectly synchronized with the task. One participant {(P23)} mentioned, ``Audio feedback was nice because it provided instant, delay-free responses. Performing tasks with audio in the background worked really well, allowing full focus on the task while receiving feedback.'' The learning curve for Audio feedback was less steep than for Haptic feedback, and musically inclined participants found it intuitive and effective. 

\textbf{Visual feedback} was the least engaging, with 20 out of 34 participants finding it distracting as it required balancing focus between the task and visual cues. While some participants appreciated its clarity and immediacy in indicating synchronization status during simple tasks, Visual feedback may have limited effectiveness for more complex tasks requiring sustained attention.  One participant {(P4)} mentioned, ``When I really focus on the task, the Visual feedback fades into the background, and I stop seeing them.'' As the tasks we presented required full visual attention, relying on Visual feedback was challenging for some.

\textbf{Haptic feedback} received mixed reactions. Initially calming and intuitive, its effectiveness diminished as tasks progressed, with some users finding it distracting. Three participants mentioned that while Haptic feedback initially felt synchronized, it became distracting over time. One participant {(P7)} noted, ``Initially haptic felt nice and felt more sync, but later on, I was paying too much attention to haptic so it could have been distracting.'' The learning curve for Haptic feedback was significant, requiring time to interpret tactile sensations correctly. Additionally, delayed feedback reduced its perceived utility. Despite these challenges, some users found Haptic feedback to be the least stressful and most intimate, suggesting its potential in tasks requiring emotional understanding and empathy.

%Participants exhibited diverse preferences for sensory feedback, influenced by their personal backgrounds and situational contexts. While some participants favored haptic responses, others found audio or visual stimuli more beneficial. This diversity underscores the complexity of designing feedback systems that accommodate a broad spectrum of users. Despite varied preferences, a consistent theme emerged regarding the necessity for accurate and synchronized feedback mechanisms to enhance task performance and collaboration. Participants emphasized the importance of timely and precise feedback to maintain synchronization with their partners during collaborative tasks. This was highlighted by 15 out of 34 participants in our exit interviews, emphasizing the critical role of feedback timing.

Participants exhibited \textbf{varied preferences} for sensory feedback based on personal backgrounds and task types, highlighting the complexity of designing universally effective systems. While some preferred Haptic feedback, others found audio or visual stimuli more beneficial, influenced by their personal and professional experiences—musicians favored Audio feedback and those with strong visual-spatial skills preferred Visual feedback. The nature of the task also mattered; Audio feedback supported rapid synchronization, Visual feedback suited simpler tasks, and Haptic feedback was considered useful for high-stress situations. Despite these differences, a key insight was the need for accurate and synchronized feedback to enhance performance and collaboration, as emphasized by 15 of 34 participants. This underscores the importance of customizable feedback systems to meet diverse user needs and task demands.

\textbf{In summary}, Audio feedback is generally preferred, Visual feedback is responsive but often goes unnoticed, and Haptic feedback, though calming, may cause fatigue over time and could be delayed.

\subsubsection{Collaboration Task and Trust Score}
Five questions determining the trust score, collected from an intertrial questionnaire, revealed that Audio feedback was perceived as the most trustworthy compared to the other two feedback methods during a collaborative task. Interestingly, in both tasks (Cup \& String session and Hand Mimicry session), the trust score rankings were identical: audio first, haptic second, and visual last. The combined mean average scores for the tasks, gathered from 68 intertrial surveys for each modality, were 4.85 (SD = 0.31) for Audio, 4.82 (SD = 0.28) for Haptic, and 4.75 (SD = 0.26) for Visual, reflecting trends noted during participant interviews. Participant {(P22)} noted that ``Haptic feedback correctly indicated how synchronized we were, which gave us confidence.'' However, some mentioned that ``audio often indicated we were out of sync when we were actually in sync, which bothered me.'' Haptic feedback was described as calming and the least stressful to use, fostering trust between participants. One participant {(P8)} stated, ``Haptic feedback helped me focus more on observing my partner, which was difficult when managing both the task and the other person, but ultimately helped with synchronization.'' Another participant {(P16)} commented, ``Haptic feedback was the only one that could reliably indicate whether or not we were synchronized with the other person.''

Despite some confusion, one participant {(P6)} said, ``More often than not, I questioned why these Audio feedback mechanisms were telling me we were not synced when we seemed pretty well synchronized with the task. This reduced my confidence in how synchronized we were.'' However, another {(P11)} noted, ``Sound helped me synchronize with the other person and best represented my status.''

% %%%%%%%%%%%%%%%%%%%%%%
% Did PAIRS who preferred modality actually have better IBS in those modality COMPARED to PAIR who didn't prefer?

\section{Discussions}
% Zoom 
% Signal noise theory -> gesture suppress Haptic feedback
% Audio -> average guy. Audio design can make a difference
% Visual + Haptic interfere task, attention
% Humans are most sensitive to auditory change -> Webber's fraction
% 

% The proposed IBS feedback system design worked for some and not for others, but it is clear that the design of the feedback can greatly impact the trust between participants and may influence their performance.

%In evaluating the effectiveness of different feedback modalities, each demonstrated distinct advantages and limitations. 
%Visual feedback, while being the least favored and statistically insignificant, highlighted issues with attention division and task integration. 
%Haptic feedback showed marginal significance and varied based on individual preferences, with effectiveness influenced by joint task duration and intensity calibration. 
%In contrast, Audio feedback emerged as the most consistently impactful modality that balances attention requirements and intrusiveness, though its efficacy was affected by subtle changes and individual auditory sensitivity. 

This discussion explores the implications of the findings for the design and optimization of feedback systems tailored to user needs and task contexts. 
Our results allow us to test our hypotheses, as shown in Table~\ref{Tab:hypothesis}. Specifically, (H1) Visual feedback did not increases the IBS level, while (H2) Auditory and (H3) Haptic feedback showed a significant positive influence on the IBS metric. Therefore, we confirmed our approach -- the external feedback in general can further augment the IBS (H4).

\begin{table}[H]
\captionof{table}{The list of the hypothesis and results}
\begin{tabular}{|c|l|c|}
\hline
   & \multicolumn{1}{c|}{Hypothesis}          & Result \\ \Xhline{1.5pt}
    H1 & Visual feedback increases IBS level   & False  \\ \hline
    H2 & Auditory feedback increases IBS level & True   \\ \hline
    H3 & Haptic feedback increases IBS level   & True   \\ \hline
    H4 & External feedback can further augments IBS & True   \\ \hline
\end{tabular}
\label{Tab:hypothesis}
\end{table}

\subsection{Visual Feedback: Distributed Attention}
% Results
Visual feedback was not statistically significant and was the least preferred modality. 
% - distributed attention, feedback-task conflict
There seemed to be a certain cognitive load applied to participants. For instance, the participants had to pay attention to both the ring representation and their own hands while performing the task~\cite{kreyenmeier_theory_2020}.
This was evident from participant comments about needing to divide their visual attention between the motor task and the feedback (P4, P8, P14, P33). This observation aligns with research suggesting that performance decrements during dual-tasking occur due to interference between tasks that rely on the same cortical regions~\cite{moisala_brain_2015}.
% Latency
There were also comments regarding the visual feedback and hand movements not matching precisely enough to be effective. This was likely due to the spatial incongruency of the feedback and the task, resulting in frequent eye movement. This points to future work on the visual feedback design that is less cognitively demanding and easier to notice (e.g., changing the color of the cup itself, or designing visual effects that are more obvious, such as changes in room brightness or color).

%some latency in accumulating the CCorr values of their actions that they believed to be in sync or out of sync and the visual feedback projections ~\cite{}.
%~\cite{moisala_brain_2015}

% \subsection{Visual}
% - Why visual doesnt work? Distraction from the task (required too much visual attention)
% - If the task wasn't visually-focused it might have work better
% - The step changes were deemed by some as difficult to notice, but extreme were considered easy to notice.. so maybe better visualization -- as noticed by Pxxx color change might have been easier to notice

%% intrinsic impact of the tactile feedback
% Need better connection to make this claim -- use citation for personalized/ calibrated haptic feedback
% \todo{Snehesh}

\subsection{Audio Feedback: Sensitivity to the Frequency Changes}
% Result
The audio feedback had the most consistent and significant impact compared to the no feedback, both statistically and practically. 
% Balanced
Compared to the visual and haptic feedback, the audio feedback emerged as a balanced option for participants, requiring less attention than the visual feedback while being less intrusive than the haptic feedback. 
This can be attributed to the fact that the joint tasks did not conflict with the auditory modality (P19, P21, P22, P26); for instance, the haptic feedback interfered with sensory experience during the task, while the visual feedback was disruptive.

% Sensitive to Sound freq changes
It is assumed that the auditory feedback effectively provided a sense of IBS levels. This aligns with Weber's Law and JND of audible frequency~\cite{houtsma_pitch_1990} -- human perception shows higher sensitivity to the audible frequency's change~\cite{merchel_psychophysical_2020} compared to haptic stimuli.
A few participants also expressed a preference for more distinguishable and pronounced audio feedback (P25, P26). This highlights the potential for further calibration and optimization of the frequency changes, which would be beneficial for a broader audience.
%and could be advantageous in fostering IBS. 
%The difficulty in learning to identify sync versus non-sync cues suggests that the audio changes may not have been effectively conveyed or perceived. 

% ?
%Several participants reported that they could not perceive the changes in auditory feedback (P18, P21). %This aligns with Weber's Law, which suggests that as the intensity of a stimulus increases, the ability to detect small changes in that stimulus diminishes ~\cite{zeng_unified_2020}. In the case of auditory feedback, when sounds were either too loud or too soft, participants may have struggled to perceive subtle variations, as their sensitivity to changes in sound was reduced. This phenomenon highlights the importance of carefully calibrating Audio feedback systems to ensure that changes in auditory cues remain perceptible, especially in environments where sound levels may vary significantly.
% 
%The inability of some participants to recognize changes in the auditory feedback may be due to the subtlety of the differences, which only individuals with musical training could discern. Several participants mentioned that the Audio feedback sounded the same to them, or they struggled to distinguish between synchronized and non-synchronized audio cues {(P18, P21)}. This indicates that the feedback might not have been distinct enough for clear recognition, particularly for those without specialized auditory skills. 

\subsection{Haptic Feedback: Aligned Attention and Calibration}
% short results
Quantitatively, the haptic feedback was marginally significant compared to no feedback.
% intensity preference
However, it is noted that the haptic feedback was also the one that most (N=13) and the least (N=14) favored. As highlighted in user preferred haptic experience~\cite{kim_defining_2020}, participants who favored were not phased by the maximum intensity we set for the experiment, while those who did not found the intensity to be distracting. This suggests a need to personalize the haptic feedback characteristics that we can control~\cite{shrestha_hdesigner_2023, pezent_syntacts_2021}, i.e. in our case intensity and the pattern design~\cite{kim_defining_2020, roy_towards_2025, whitmore_improving_2024}.
% that the haptic feedback was the affected by individual preferences and sensitivity to the haptic stimuli.

% not favor
In fact, those who did not favor the haptic feedback (P5, P6, P13, P14) mentioned that the changes were difficult to notice or the feedback was too ``intrusive'' or ``obstructive.'' Additionally, the effectiveness of the haptic feedback also depended on the task duration, as participants became desensitized to the feedback over time~\cite{graczyk_sensory_2018}, causing its impact to diminish.

% favor
Participants who favored the haptic feedback (P7, P18, P20, P23, P24) noted that they benefited from not needing to divide their attention and be able to feel the state of their sync~\cite{pasquero_haptic_2011}. This aligns with the comments in the visual feedback regarding the lower efficacy due to the distributed attention.

\subsection{Feedback Preference and Synchrony}
Out of 16 pairs, 10 pairs shared the same top preference for the type of feedback. Notably, these pairs demonstrated stronger synchrony compared to those with differing preferences. On average, the CCorr values for pairs with the same preferred modality increased by 14\% in that modality compared to No Feedback. Conversely, pairs with mismatched preferences or those who had the least preference saw little to no improvement or even a decrease in CCorr values compared to the No Feedback condition. 

%This suggests the potential for a feedback system design that aligns with each pair's preferences to enhance synchrony and overall effectiveness.

The results indicated the potential for designing a feedback system tailored to each pair's preferences to enhance synchrony and overall effectiveness. However, considering that excessively distinct stimuli can degrade synchrony, a future study to validate this approach and explore the strategy may be valuable.

\section{Design Implications and Potential Applications}
\label{application}
\subsection{Design Implications with Haptic and Auditory IBS Feedback}
% introduce further use cases 
%Our findings suggest several practical applications for advanced feedback systems designed to enhance communication and synchronization across various settings.
% noisy environment, audio sensitive task, - haptic 
% - audio

Drawing from our quantitative and qualitative findings, we synthesize key design implications to improve IBS.

% Audio Guideline
\textbf{Auditory IBS feedback} based on the frequency changes is an optimal choice for general tasks, including motor tasks, due to its minimal interference.
However, it presents challenges, as it is ineffective in noisy environments or situations requiring audio sensitivity, such as public spaces, playing musical instruments, or engaging in conversations.

% Haptic Guideline
For such cases, the \textbf{haptic IBS feedback} based on the rhythmic vibrotactile feedback would be an optimal choice since it does not exhibit any sound and is not affected by surrounding noise. Additionally, it does not expose the IBS levels publicly, preserving users' privacy and conversations that might be hindered by the use of headphones.
On the other hand, a task should be carefully chosen as it might create conflicts in case the task heavily relies on the sensory inputs of the human fingers (e.g. construction, dexterous crafting activity, texture recognition).

\subsection{Potential Applications}
Here we illustrate two potential applications based on the design implications:

% Scenario for audio
% motor task, rehabilitation
In the scenario of \textbf{physical therapy}, it is beneficial to strengthen the bond and sympathy between a therapist and a patient~\cite{miskovic_changes_2011} for better communication during a motor task.
Our \textit{auditory IBS feedback} would be useful for this situation, as the patient can focus on the rehabilitation task while their IBS is enhanced, potentially improving the overall therapeutic experience including motivation, comfort, sympathy, and a sense of togetherness.

% Scenario for Haptic
% Music, public, privacy, brainstorming
In the scenario of \textbf{brainstorming}, it is crucial to maintain effective team and verbal communication when forming creative ideas. Our \textit{haptic IBS feedback} would be an optimal choice for this case, potentially enhancing their sense of closeness~\cite{liu_team-work_2021}, trust~\cite{liu_interactive_2018}, and agreeableness~\cite{zhang_group_2021}, while not interfering with their conversations. This could be extended to share experiences and foster a sense of togetherness across distances. By streaming feedback through digital platforms (e.g. a video conference), audiences and collaborators who are not physically present can be immersed in the shared experience, therefore enhancing engagement and interaction.

\section{Limitations and Future Work}
% task adaptive design / personalization / Multimodal approach
% fidelity of the feedback design -> limitation
The proposed feedback method shows its limitations and further room for improvement in the following criteria: task adaptive design, personalization, multimodal approach, and signal noise reduction.

\textbf{First, task-specific adjustments and feedback mechanisms may optimize IBS enhancement.} For instance, Visual feedback should account for field of view and focal attention while accommodating the task. Given the study's short task duration (1-2 minutes), participants had to split their attention, reducing the feedback's impact. Future research should test longer collaboration sessions, such as in classrooms, to mitigate these constraints and strengthen feedback effects.

\textbf{Secondly, the effect of personalized feedback needs to be explored.} For instance, perception of the auditory feedback depends on factors like familiarity, cultural background, and musical expertise~\cite{di_stefano_consonance_2022}. P25 and P26 expressed interest in adjusting feedback to their preference. Collecting user preferences across diverse backgrounds and contexts could help researchers design systems that customize feedback, such as tones, vibration patterns, intensities, or ring sizes, to better suit users. Future work could develop a platform enabling pairs to modify these settings collaboratively. However, excessive personalization may have unintended consequences. If participants in a pair receive highly divergent stimuli, this could increase IBS uncertainty. A potential solution is providing shared choices that balance preferences, such as lowering haptic intensity for the pair if one participant finds it too strong, which may enhance IBS overall. Still, further research is needed to validate whether personalized feedback increases synchrony and to develop effective personalization strategies.

%if one of the participant is not musically trained, instead of musical notes, using metronome or, loudness in audio that are adjusted to levels comfortable to the pair. 

\textbf{Thirdly, testing more feedback parameters and patterns}, such as frequency, strength, and/or pattern of the feedback, in each feedback design would lead to distinguishing the intrinsic reaction of the external modality from the proposed IBS enhancement. This would present a better understanding of the relationship between the intrinsic property of external stimuli and IBS, ultimately leading to a higher prosocial impact. This can be expanded to combining multiple modalities that could compensate for the insufficient estimations gained by one modality~\cite{sigrist_augmented_2013}.

\textbf{Finally, minimizing EEG signal noise in real-time} is crucial for responsive feedback systems. Removing entire EEG signal epochs due to motion artifacts caused latency and misalignment with tasks and feedback. Advanced noise reduction algorithms or noise-robust EEG headsets with improved hardware could enhance signal quality and system responsiveness. However, EEG devices can be intrusive, and social acceptability remains a challenge. Improving comfort and analysis resolution will be essential for real-world applications.

In our study, we aimed to maintain consistent fidelity across feedback methods to observe the characteristics of each modality's impact on IBS. However, this approach meant that we explored general uses of visual, audio, and haptic feedback, rather than task-specific methods such as augmented reality spatial UI or specialized visual and audio feedback for particular tasks. Future research could focus on these task-specific feedback methods to further enhance IBS.

\section{Conclusions} 
This study investigated how Visual, Auditory, and Haptic feedback influence Inter-brain Synchronization (IBS) and provided design recommendations for feedback systems aimed at enhancing IBS. By employing EEG to measure IBS metrics, our results reveal that auditory feedback had the most substantial and consistent impact on improving IBS. Haptic feedback also positively affected IBS intensity, whereas Visual feedback, delivered through on-body projection mapping, did not show significant improvements compared to no feedback. These findings validate our hypothesis that feedback generally enhances IBS intensity and highlight opportunities for further improvement through task-adaptive design, personalization, and multimodal approaches. This research offers crucial insights for developing effective feedback systems that can boost IBS and enhance user experiences.

\begin{acks}
This work was supported by the JST Moonshot R{\&}D Program (JPMJMS2013), UMD's Immersive Media Design program, and ArtsAMP grant. Any opinions, findings, and conclusions, or recommendations expressed in this material are those of the authors and do not necessarily reflect the views of any funding agencies. We are grateful for the feedback from our reviewers.
\end{acks}

\bibliographystyle{ACM-Reference-Format}
\bibliography{yujnkm.bib}
\end{document}